\documentclass[traditabstract]{aa} 
\usepackage{graphicx}
\usepackage{txfonts}
\usepackage{natbib}
\begin{document}
\title{The distance to the young open cluster Westerlund 2\thanks{Based on observations carried out at Las Campanas Observatory},\thanks{Full photometric data 
are available at the CDS via anomymous ftp to cdsarc.u-strasbg.fr
(130.79.128.5) or via http://cdsarc.u-strasbg.fr/viz-bin/qcat?J/A+A/999/A999}}
 
\titlerunning{The young open cluster Westerlund 2}

\author{Giovanni Carraro\thanks{On leave from Dipartimento di Fisica e Astronomia, Universit\'a di Padova, Italy} 
     \inst{1}, David Turner\inst{2}, Daniel Majaess\inst{2}, Gustavo Baume\inst{3}
     }

   \institute{ESO, Alonso de Cordova 3107, 19001,
      Santiago de Chile, Chile\\
       \email{gcarraro@eso.org}
     \and
      Department of Astronomy and Physics, Saint Mary's University,
        Halifax, NS B3H 3C3, Canada\\
        \email{turner,dmajaess@ap.smu.ca}
       \and
      Facultad de Ciencias Astron\'omicas y Geof\'isicas (UNLP), Instituto
        de Astrof\'isica de La Plata (CONICETUNLP), Paseo del Bosque s/n, La
        Plata, Argentina\\
       \email{gbaume@fcaglp.unlp.edu.ar}
       }
\authorrunning{Carraro et al.}

\date{Received March , 2013; accepted , 2013}

\abstract
{A new X-ray, {\it UBVR}$I_c$, and {\it JHK$s$} study of the young cluster Westerlund~2 was undertaken to resolve discrepancies tied to the cluster's distance. Existing spectroscopic observations for bright cluster members and new multi-band photometry imply a reddening relation towards Westerlund~2 described by $E_{U-B}/E_{B-V}=0.63 + 0.02\;E_{B-V}$. Variable-extinction analyses for Westerlund~2 and nearby IC~2581 based upon spectroscopic distance moduli and ZAMS fitting yield values of $R_V=A_V/E_{B-V}=3.88\pm0.18$ and $3.77\pm0.19$, respectively, and confirm prior assertions that anomalous interstellar extinction is widespread throughout Carina (e.g., Turner 2012). The results were confirmed by applying the color difference method to {\it UBVR$I_c$JH$K_s$} data for 19 spectroscopically-observed cluster members, yielding $R_V=3.85\pm0.07$. The derived distance to Westerlund~2 of $d=2.85\pm0.43$ kpc places the cluster on the far side of the Carina spiral arm. The cluster's age is no more than $\tau \sim2\times10^6$ yr as inferred from the cluster's brightest stars and an X-ray (Chandra) cleaned analysis of its pre-main-sequence demographic. Four Wolf-Rayet stars in the cluster core and surrounding corona (WR20a, WR20b, WR20c, and WR20aa) are likely cluster members, and their inferred luminosities are consistent with those of other late-WN stars in open clusters. The color-magnitude diagram for Westerlund~2 also displays a gap at spectral type B0.5~V with associated color spread at higher and lower absolute magnitudes that might be linked to close binary mergers. Such features, in conjunction with the evidence for mass loss from the WR stars, may help to explain the high flux of $\gamma$ rays, cosmic rays, and X-rays from the direction towards Westerlund~2.}


\maketitle
  
\section{Introduction}

Westerlund 2 is a compact young open cluster embedded in the H~{\sc ii} region RCW~49. The cluster lies in the direction of the Carina spiral arm ($\ell,b=284^{\circ}.4,-0^{\circ}.34$), and descriptions of prior studies of the object and region are provided by \citet{as07} and \citet{va13}. Most noteworthy is that the distance to Westerlund~2 has been the subject of lively debate \citep[e.g.,][]{ra07,as07}, with estimates ranging from 2 kpc to more than 8 kpc from the Sun. That range propagates into uncertainties in the size, mass, and luminosity determinations for the cluster. Consequently, establishing a reliable cluster distance is vital, and depends directly on how interstellar extinction is treated because of the large reddening for Westerlund~2 members. Such an analysis is particularly important given the cluster's location in the vicinity of the Great Carina Nebula and complex \citep[e.g.,][]{ca04b}, where anomalous dust extinction ($R_V=A_V/E_{B-V} \simeq 4$) appears widespread \citep[][see also \citealt{ba09}]{tu12}. 

Westerlund~2 is also located in the direction of one of the Galaxy's strongest sources of $\gamma$-rays \citep{ah07,ab11}, and is also coincident with a strong source of X-rays \citep{ts07}, which suggests that the physical conditions necessary for producing such sources of radiation must be present in the cluster \citep[see][]{be07} if a physical association with Westerlund~2 is confirmed. Several Wolf-Rayet stars in the cluster and surroundings \citep*{mo91,rl11} are potential sources of high energy radiation, as, for example, is the case for the northern hemisphere cluster Berkeley~87 \citep{ma96,tl01,ab09}. Berkeley 87 is spatially coincident with the strongest northern hemisphere source of $\gamma$ rays, and is a cluster that contains the WO2 Wolf-Rayet star WR~142 and several unusual high-mass objects \citep{tf82}. Strong mass loss has been demonstrated to be a feature of luminous members of Berkeley~87 \citep{te10}, and may also be the case for luminous members of Westerlund~2. Such characteristics are important considerations for identifying potential sources of high energy radiation from the clusters \citep{be07}, in addition to the observed frequency dependence of radiation from them.

An important consideration in all such studies is to establish the correct distance to Westerlund~2. The present analysis employs new $UBV RI_c$ photometry 
(see also Carraro et al. 2012, in press) to tackle the cluster distance through zero-age main sequence (ZAMS) fitting, including a new analysis of the reddening law in the region. The revised results have ramifications pertinent to the determination of the cluster mass and the reputed membership of the Wolf-Rayet stars WR20aa and WR20c \citep*{rl11}.

\begin{figure*}
\begin{center}
\includegraphics[width=14cm]{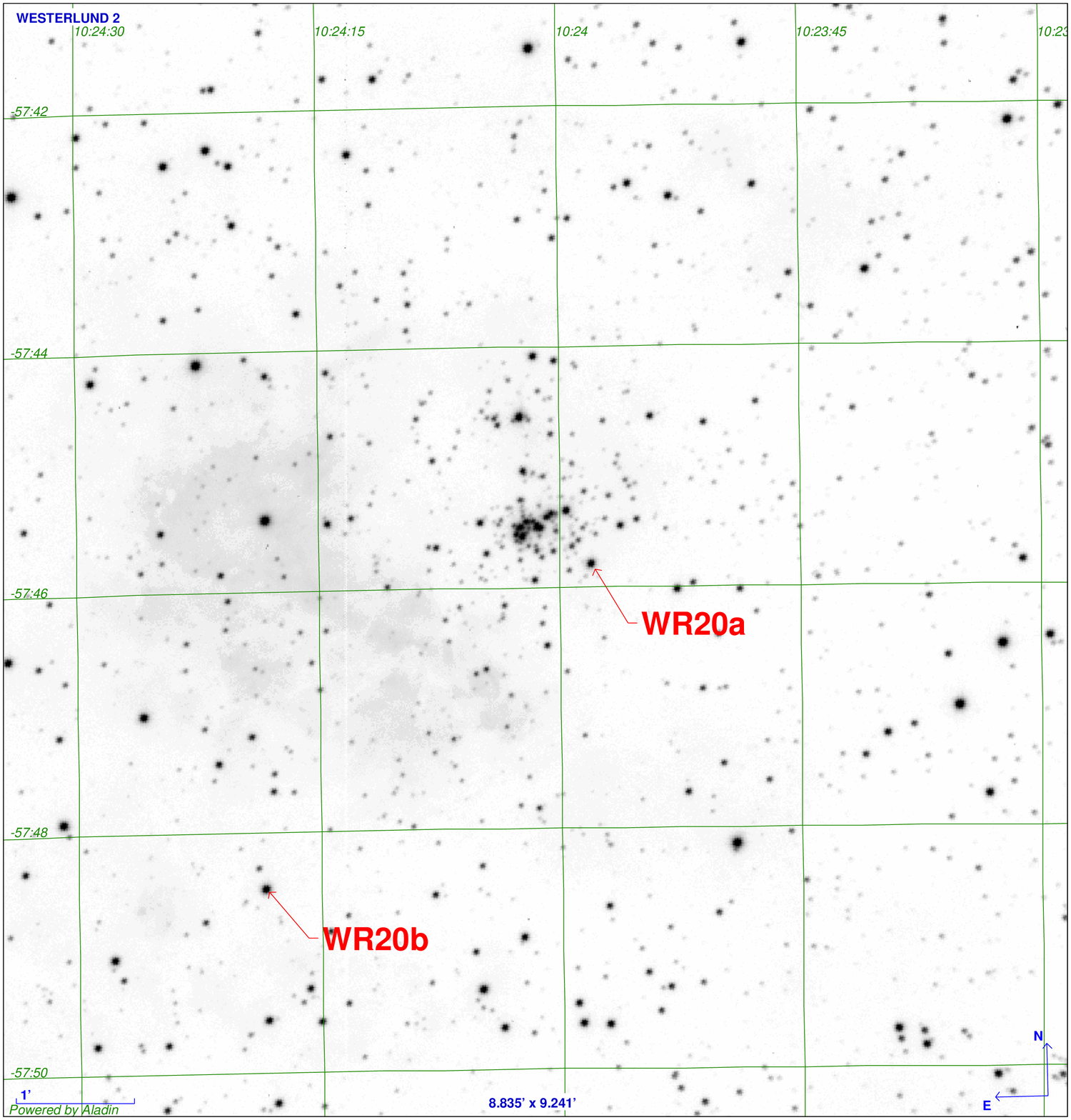}
\end{center}
\caption{A 40 s exposure in the {\it R} band showing a portion of the
field studied here. To highlight the cluster core the original image
was cropped to a $9\arcmin \times 9\arcmin$ field. Two important WR
stars discussed here are highlighted in the figure.}
\label{fig1}
\end{figure*}

\section{Observations and data reduction}
\label{s-ob}

Westerlund 2 was originally observed at the Cerro Tololo Inter-American Observatory on the nights of March 10 and 13, 2010, using the 1-m ex-YALO telescope, operated by the SMARTS consortium.\footnote{\tt http://http://www.astro.yale.edu/smarts} The camera is equipped with an STA 4064$\times$4064 CCD\footnote{\texttt{http://www.astronomy.ohio-state.edu/Y4KCam/ detector.html}} with 15-$\mu$m pixels, yielding a scale of $0.289 \arcsec$/pixel and a field-of-view (FOV) of $20^{\prime} \times 20^{\prime}$ at the Cassegrain focus of the CTIO 1-m telescope. The CCD was operated without binning at a nominal gain of 1.44 e$^-$/ADU, implying a readout noise of 7~e$^-$ per quadrant (the detector is read by means of four different amplifiers).

All observations were carried out in good seeing conditions (always less than $1.3 \arcsec$), but not photometric conditions. Our {\it UBV} instrumental photometric system was defined by the use of a standard broad-band Kitt Peak {\it UBV} set of filters.\footnote{\texttt{http://www.astronomy.ohio-state.edu/Y4KCam/ filters.html}}
The set of images obtained is listed in Table~\ref{tbl1}. 

\setcounter{table}{0}
\begin{table}
\caption[]{{\it UBV$I_c$} photometric observations of Westerlund~2 obtained at the CTIO in March 2010.}
\label{tbl1}
\begin{center}
\begin{tabular}{@{\extracolsep{-0pt}}lccc}
\hline \hline
\noalign{\smallskip}
Date & Filter & Exposure (s) & airmass\\
\noalign{\smallskip}
\hline
\noalign{\smallskip}
2010-03-10 &\textit{U} & 30, 3$\times$200   & 1.19--1.21\\
      &\textit{B} & 20, 150    & 1.15$-$1.15\\
      &\textit{V} & 20, 100    & 1.14$-$1.14\\
      &\textit{$I_c$} & 3$\times$20, 2$\times$100  & 1.16--1.17\\
\noalign{\smallskip}
\hline
\noalign{\smallskip}
2010-03-13 &\textit{U} & 30, 2000    &1.15--1.22\\
      &\textit{B} & 30, 1500    &1.16--1.23\\
      &\textit{V} & 2$\times$20, 2$\times$900  &1.17--1.28\\
\noalign{\smallskip}
\hline
\end{tabular}
\end{center}
\end{table}

In order to secure photometric calibration and add {\it R}-band and deep {\it $I_c$}-band images, the field under analysis was observed at Las Campanas Observatory (LCO) on the night of May 17, 2010, as summarized in Table~\ref{tbl2}, which lists useful details of the observations, such as filter coverage, air mass range, and exposure times and sequences. The observations made use of the SITe$\#$3 CCD detector on the Swope 1.0m telescope\footnote{\texttt {http://www.lco.cl/telescopes-information/ henrietta-swope.html}}. With its pixel scale of $0.435\arcsec$/pixel, the CCD gives a coverage of $14.8 \arcmin \times 22.8 \arcmin$ on the sky. The night of observation was photometric with seeing ranging from 0.9 to $1.4 \arcsec$ . The field observed is shown in Fig.~\ref{fig1}, where a bias- and flat-field-corrected image in the {\it R} band (40 s) is shown.

The transformation from the instrumental system to the standard Johnson-Kron-Cousins system, as well as corrections for atmospheric extinction, were determined through multiple observations of stars in Landolt's areas PG~1047, PG~1323, and MarkA \citep{la92}. Large ranges in air mass ($\sim1.06$ to $\sim1.91$) and color ($-0.5 \leq (B-V) \leq 1.5$) were covered.

\setcounter{table}{1}
\begin{table}
\caption[]{{\it UBVR$I_c$} photometric observations of Westerlund~2 and standard star fields at LCO in May 2010.}
\label{tbl2}
\begin{center}
\begin{tabular}{@{\extracolsep{-0pt}}lccc}
\hline \hline
\noalign{\smallskip}
Target & Filter & Exposure (s) & airmass\\
\noalign{\smallskip}
\hline
\noalign{\smallskip}
MarkA    & \textit{U} & 2$\times$240        &1.07--1.38\\
       & \textit{B} & 2$\times$180        &1.07--1.41\\
       & \textit{V} & 2$\times$60         &1.07--1.38\\
       & \textit{R} & 40, 50        &1.06--1.35\\
       & \textit{$I_c$} & 40, 50        &1.06--1.34\\
PG~1323   & \textit{U} & 180         &1.11\\
       & \textit{B} & 120         &1.12\\
       & \textit{V} & 30          &1.13\\
       & \textit{R} & 30          &1.13\\
       & \textit{$I_c$} & 30          &1.13\\
PG~1047   & \textit{U} & 20, 180       &1.14--1.91\\
       & \textit{B} & 15, 180       &1.14--1.88\\
       & \textit{V} & 2$\times$10, 30       &1.15--1.83\\
       & \textit{R} & 10, 30, 40       &1.14--1.81\\
       & \textit{$I_c$} & 410, 30        &1.15--1.86\\
Westerlund~2 & \textit{U} & 5, 300, 2000        &1.15--1.26\\
       & \textit{B} & 5, 120, 1800        &1.15--1.25\\
       & \textit{V} & 3, 30 , 900        &1.15--1.27\\
       & \textit{R} & 3, 40, 600      &1.16--1.26\\
       & \textit{$I_c$} & 3, 30, 900      &1.18--1.29\\
\noalign{\smallskip}
\hline
\end{tabular}
\end{center}
\end{table}

\subsection{Photometric reductions}

Basic calibration of the CCD frames was done using the IRAF\footnote{IRAF is distributed by the National Optical Astronomy Observatory, which is operated by the Association of Universities for Research in Astronomy, Inc., under cooperative agreement with the National Science Foundation.} package CCDRED. Zero exposure frames and twilight sky flats were taken every night for standardization purposes. All frames were pre-reduced applying trimming, bias, and flat field correction. Prior to flat-fielding, all LCO frames were corrected for linearity,
following the recipe discussed in \citet{ha06}.

Photometry was then performed using the IRAF DAOPHOT/ALLSTAR and PHOTCAL packages. Instrumental magnitudes were extracted following the point-spread function (PSF) method \citep{st87}. A quadratic, spatially variable, master PSF (PENNY function) was adopted because of the large field of view of the detector. Aperture corrections were then determined using aperture photometry of a suitable number (typically 15 to 20) of bright, isolated, stars across the field. The corrections were found to be stable across the field, and vary from 0.130 to 0.270 mag, depending upon the filter. Lastly, the PSF photometry was aperture corrected, filter by filter.

\subsection{Photometric calibration}

After the removal of problematic stars and stars having only a few observations in Landolt's catalog \citep{la92}, a photometric solution for the run was extracted from
a total of 43 measurements per filter. The resulting calibration relations are (see also Fig.~2):\\

\noindent
$ U = u + (4.902\pm0.010) + (0.41\pm0.01) \times X + (0.129\pm0.020) \times (U-B)$ \\
$ B = b + (3.186\pm0.012) + (0.31\pm0.01) \times X + (0.057\pm0.008) \times (B-V)$ \\
$ V = v + (3.115\pm0.007) + (0.17\pm0.01) \times X - (0.057\pm0.011) \times (B-V)$ \\
$ R = r + (2.986\pm0.008) + (0.11\pm0.01) \times X - (0.039\pm0.011) \times (V-R)$ \\
$ I = i + (3.426\pm0.011) + (0.07\pm0.01) \times X + (0.091\pm0.012) \times (V-I)$ \\

The final rms residuals of the fitting in the present case are $\pm0.025$, $\pm0.020$, $\pm0.013$, $\pm0.018$, and $\pm0.013$ in {\it U, B, V, R}, and {\it $I_c$}, respectively.

Global photometric uncertainties were estimated using the scheme developed by \citet[][Appendix A1]{pa01}, which takes into account the uncertainties resulting from the PSF fitting procedure (i.e., from ALLSTAR), and the calibration uncertainties corresponding to the zero point, color terms, and extinction corrections. The final catalogue contains 3481 \textit{UBVR$I_c$} and 12879 \textit{V$I_c$} entries.

\subsection{Completeness and astrometry}

Completeness corrections were determined by running artificial star experiments on the data. Basically, several simulated images were created by adding artificial stars to the original frames. The artificial stars were added at random positions and had the same color and luminosity distribution as the sample of true stars. To avoid potential overcrowding, up to 20\% of the original number of stars were added in each simulation. Depending on the frame, between 1000 and 5000 stars were added in such fashion. The results are summarized in Table~\ref{comp}.

\begin{table}
\caption{Completeness study as a function of the filter.}
\begin{center}
\begin{tabular}{lr r r r r }
\hline\hline
$\Delta$ Mag  &  U & B & V & R & I\\
\hline
12-13 &  100\% &  100\% & 100\% & 100\% &  100\% \\ 
13-14 &  100\% &  100\% & 100\% & 100\% &  100\% \\ 
14-15 &  100\% &  100\% &  100\% & 100\% &  100\% \\ 
15-16 &  100\% &  100\% &  100\% & 100\% &  100\% \\ 
16-17 &  100\% &  100\% &  100\% & 100\% &  100\% \\ 
17-18 &  100\% &  100\% &  100\% & 100\% &  100\% \\ 
18-19 &   90\% &   93\% &  100\% & 100\% &  100\% \\ 
19-20 &   73\% &   75\% &  100\% & 100\% &  100\% \\ 
20-21 &   50\% &   55\% &   95\% &  90\% &  100\% \\ 
21-22 &        &         &   67\% &  57\% &   88\% \\ 
22-23 &        &         &   50\% &        &   68\% \\ 
\hline
\end{tabular}
\end{center}
\label{comp}
\end{table}

The optical catalog was then cross-correlated with 2MASS, which resulted in a final catalog that included \textit{UBVR$I_c$} and \textit{JHK$_{s}$} photometry. As a by-product, pixel (i.e., detector) coordinates were converted to RA and DEC for equinox J2000.0, thus providing 2MASS-based astrometry useful for {\it e.g.} spectroscopic follow-up.

\subsection{Comparison with previous photometry}

Optical data for Westerlund~2 are scanty. \citet{mv75} obtained photoelectric photometry for 9 stars in the cluster, but caution readers that the region is complicated and most stars could be binaries. Photoelectric fixed-aperture photometry in crowded fields also carries the classical difficulties of handling blends of stars \citep{pa01}. CCD photometry by \citet*{mo91} extends the observed sample of cluster stars to $V \simeq 19$, with the photometry being calibrated using the \citet{mv75} observations. \citet{ca04} remarked that their $U$-band photometry was offset by upwards of $0.2$ mag relative to the \citet{mo91} data. Moffat noted in a private communication that the offset could be related to unaccounted-for color terms in the \citet{mo91} dataset. Another source for the discrepancy may be related to aperture corrections when PSF photometry is used for cluster stars and fixed aperture photometry is used for standard stars. No details about that procedure are reported by \citet{mo91}. 

An independent set of optical data has recently been published by \citet{ra07}. \citet{ra07} find that the \citet{ca04} photometry is brighter in {\it V} by $\sim0.2$ mag relative to their photometry. No mention is made by \citet{ra07} concerning the size of the aperture correction. To our knowledge, no other photometric studies have been conducted on the cluster to date.

\begin{figure}
\begin{center}
\includegraphics[width=\columnwidth]{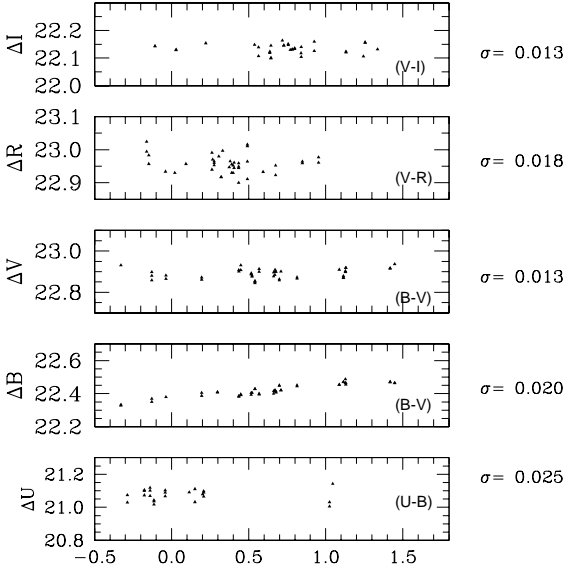}
\end{center}
\caption{Photometric solution in UBVRI for standard stars (see Table 2 for
details). $\sigma$, on the right-hand side, indicates the rms of the fit.}
\label{fig2}
\end{figure}

\begin{figure}
\begin{center}
\includegraphics[width=\columnwidth]{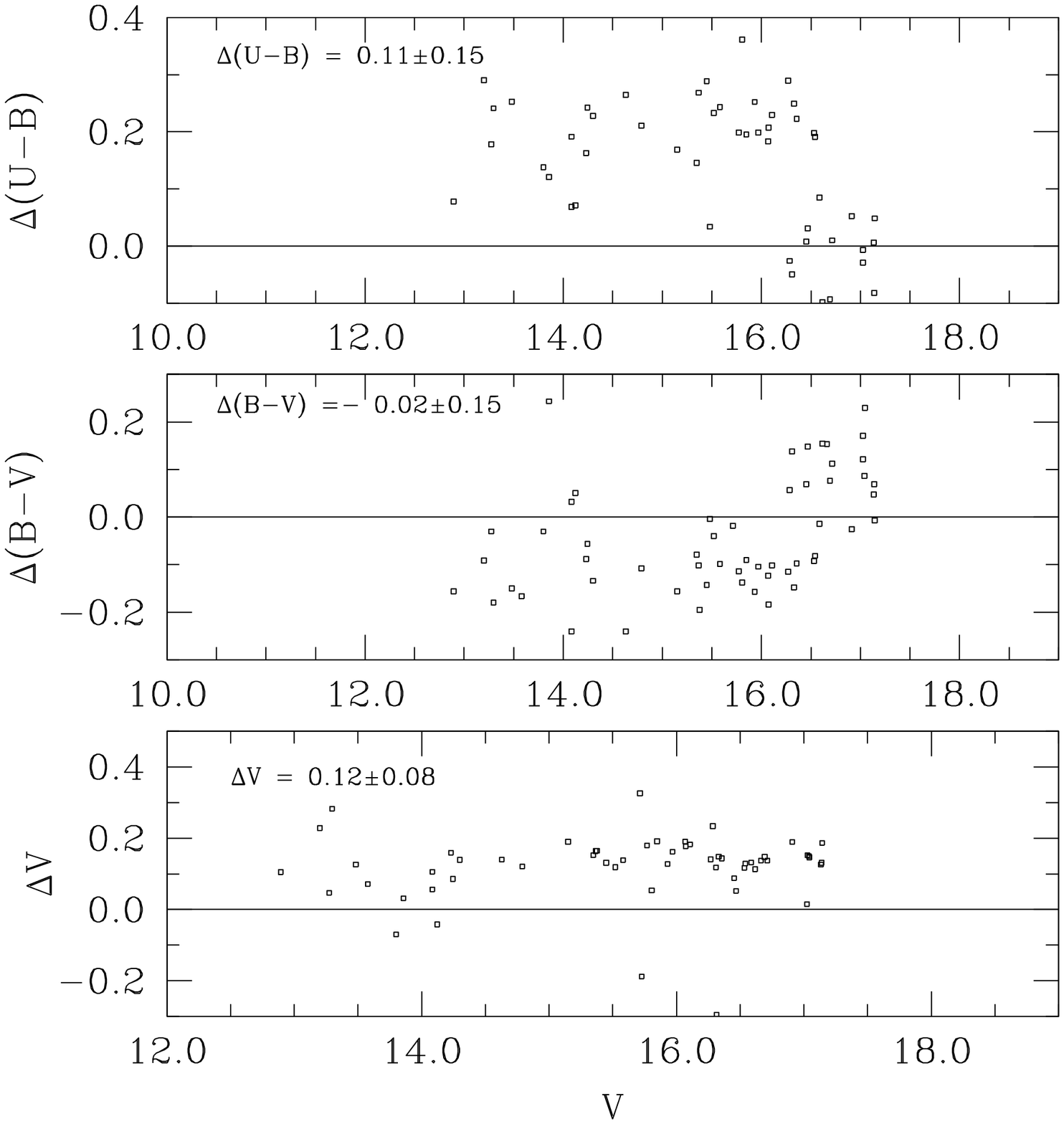}
\end{center}
\caption{Comparisons of the photometry presented in this paper with that of \citet{mo91}. }
\label{fig3}
\end{figure}

\begin{figure}
\begin{center}
\includegraphics[width=\columnwidth]{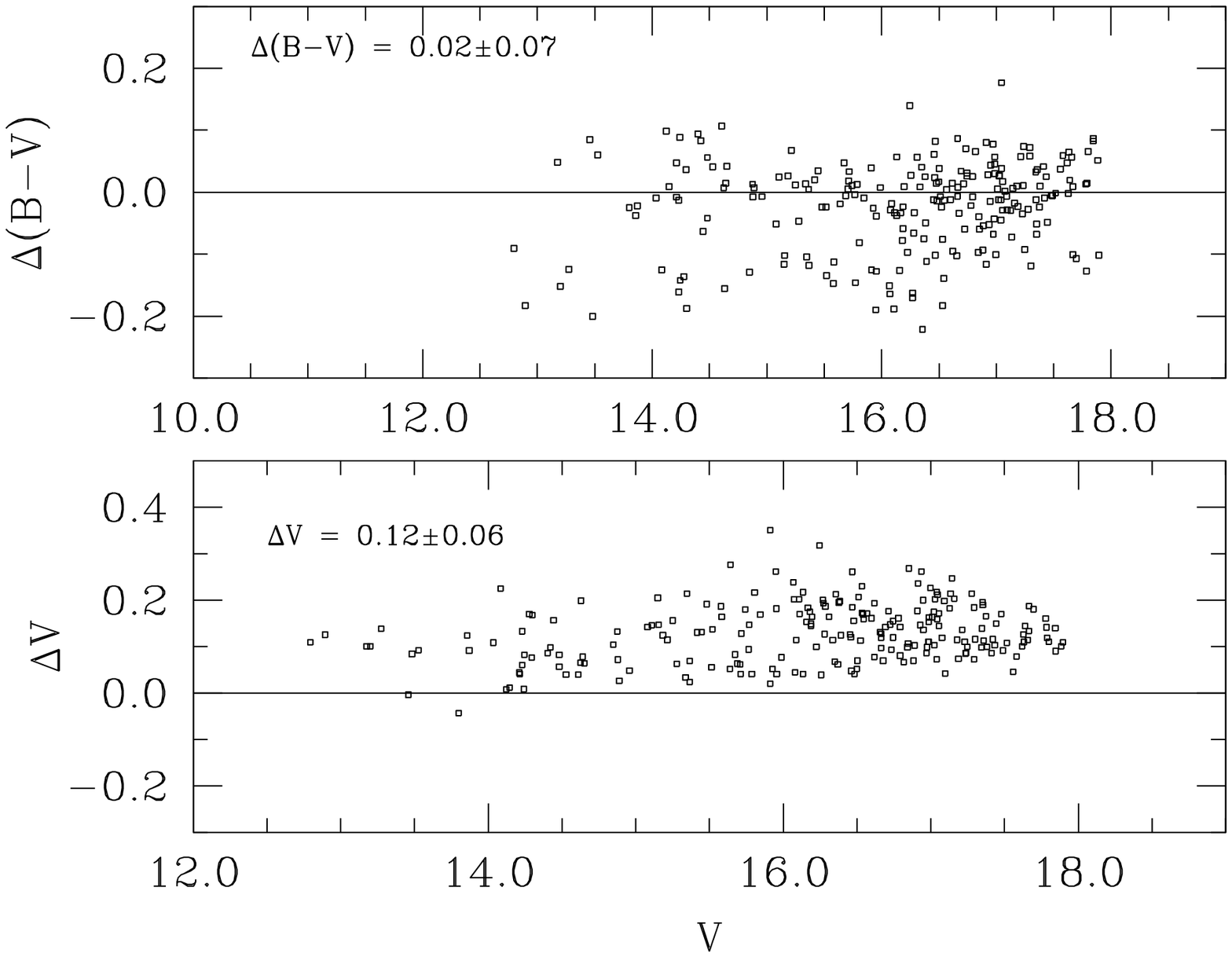}
\end{center}
\caption{Comparisons of the photometry presented in this paper with that of \citet{ra07} .}
\label{fig4}
\end{figure}

\begin{figure}
\begin{center}
\includegraphics[width=\columnwidth]{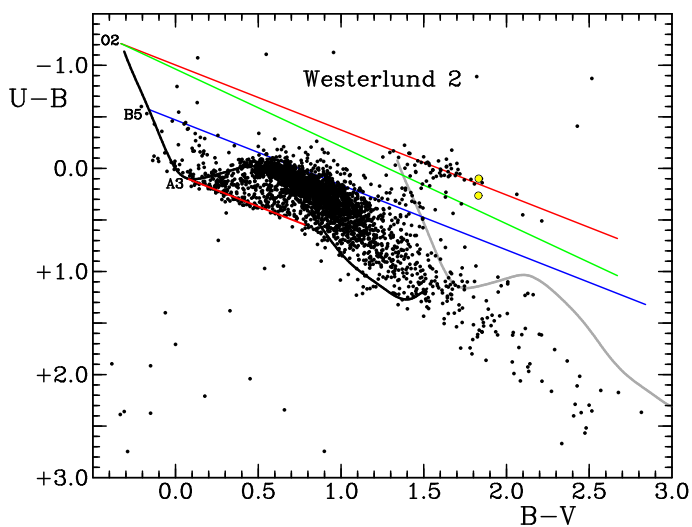}
\end{center}
\caption{Observed colors for stars in Westerlund~2, with data for the Wolf-Rayet stars WR~20a and WR~20b plotted as circled yellow points. The intrinsic relation for dwarfs is plotted as a solid black curve, with the same relation reddened by $E_{B-V}=1.65$ as a gray curve. Reddening lines of slope $E_{U-B}/E_{B-V}=0.64$ are plotted relative to the hottest O stars and A3 dwarfs (red lines), and B5 dwarfs (blue line). A similar reddening line of slope $E_{U-B}/E_{B-V}=0.75$ is shown for the hottest O stars (green line).}
\label{fig5}
\end{figure}

Since the panchromatic response of current CCD detectors is not well suited to the replication of the Johnson {\it UBV} system, particularly for the {\it U}-band, it is important to establish the general accuracy of the present observations prior to an analysis of Westerlund~2. To begin, a comparison of the new photometry presented here was made with respect to the data of \citet{mo91}. A cross-correlation of the two catalogs revealed 61 stars in common with {\it UBV} photometry. A comparison is illustrated in Fig.~\ref{fig3} , where the differences in {\it V, B--V}, and {\it U--B} are plotted in the sense Moffat et al. (M91) minus the present photometry (C13). The comparison yields offsets of:
\begin{eqnarray}
\nonumber
V_{M91} - V_{C13} &=& 0.12 \pm 0.10 \\ \nonumber
(B-V)_{M91} - (B-V)_{C13} &=& -0.02 \pm 0.15 \\ \nonumber
(U-B)_{M91} - (U-B)_{C13} &=& 0.11 \pm 0.15 \nonumber
\end{eqnarray}

\noindent
The large deviation at faint $V$ magnitudes is the result of poor detector sensitivity, which is common when comparing photometry from modern detectors to older data sets \citep{ca11}.

The study by \citet{ra07} is tied to {\it BV} observations. A cross-correlation with their photometry yields 226 stars in common, with the comparisons shown in Fig.~\ref{fig3}. The offsets are:
\begin{eqnarray}
\nonumber
V_{R07} - V_{C13} = 0.12 \pm 0.09 \\ \nonumber
(B-V)_{R07} - (B-V)_{C13} = 0.01 \pm 0.20 \nonumber
\end{eqnarray}
\noindent
The same {\it V} offset exists as in the comparison with the \citet{mo91} data, yet the {\it B--V} colors agree. 

\section{Photometric Analysis}
\label{s-pa}

The accuracy of the CCD photometry presented here was tested using the observations themselves, particularly in light of the offsets from previously published photometry described in \S2.4. The {\it UBV} data appear to be closely matched to the Johnson system, as shown in Fig.~\ref{fig5}, which is a color-color diagram for observed stars in the field of Westerlund~2. The observations are plotted relative to the intrinsic {\it UBV} relation for dwarf stars, the same relation reddened by $E_{B-V}=1.65$, and reddening lines of slope $E_{U-B}/E_{B-V}=0.64$ and $0.75$ for the hottest O-type stars, with the former also plotted relative to the intrinsic colors for B5 and A3 dwarfs. There are a number of highly deviant points in Fig.~\ref{fig5}, a situation typical of stars in crowded fields or stars near the limits of reliable photometry with the detector. Yet it is noteworthy that there are also many stars that closely match the intrinsic relation for unreddened dwarfs, which is often a good test of how closely the observations match the Johnson {\it UBV} system \citep[see][]{te12b}.

\subsection{Reddening law}

The clump of stars near ({\it B--V, U--B}) = (1.65, 0.00) represents the reddened colors of massive O-type members of Westerlund~2. The two plotted reddening lines from the earliest main sequence spectral types are consistent with the run of observed colors for the stars only for a reddening slope near $E_{U-B}/E_{B-V}=0.64$. A steeper reddening relation such as that plotted for $E_{U-B}/E_{B-V}=0.75$ does not match the observations for the O-stars; the run of colors for the stars is noticeably different and would imply sizable ultraviolet excesses for most of them, a highly improbable situation. Reddening lines of slope 0.64 are also a close match to the colors of stars reddened from the A3 ``kink'' in the intrinsic relation for dwarfs at {\it B--V} = 0.08, as well as from the intrinsic colors for B5 dwarfs, the latter appearing to form a consistent blue limit for the colors of less-reddened foreground stars in the Westerlund~2 field. If there were a color-dependent error in the {\it U--B} colors from the CCD photometry, such consistent trends would not be evident in the data.

A further test was made by examining the color excesses for spectroscopically observed members of Westerlund~2, including stars classified spectroscopically by \citet{va13}. The intrinsic colors of O-type dwarfs and giants vary little with spectral type, but for consistency we rederived spectral types for those stars in the compilation of spectra plotted by \citet{ra07}, without reference to earlier classifications, although the results are generally similar. The temperature and luminosity subtypes were tied closely to the Walborn system \citep{wa71a,wa71b,wa72,wp90}, particularly in luminosity, with reference to published spectral line criteria. Color excesses, $E_{B-V}$ and $E_{U-B}$, were derived for the stars using an unpublished set of intrinsic colors for early-type stars established by DGT \citep[see][]{te12a} through a combination of published tables by \citet{jo66} and \citet{fi70}, in conjunction with observational studies of young clusters. The \citet{va13} classifications were adopted as published. The results are summarized in Table~\ref{tbl3}, where the star numbering is from \citet{mo91}.

\begin{figure}
\begin{center}
\includegraphics[width=8cm]{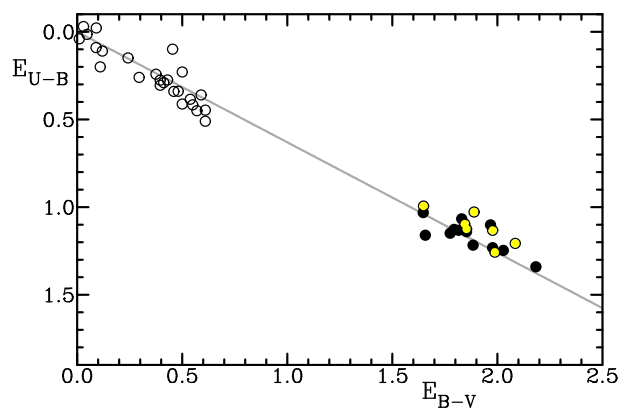}
\end{center}
\caption{A plot of {\it UBV} colour excesses for spectroscopically observed stars in Westerlund~2 (filled circles) and within $1\degr$ of it (open circles). Circled yellow points represent spectra for Westerlund~2 stars from \citet{va13}. The plotted reddening relation is described by $E_{U-B}/E_{B-V}=0.63 + 0.02\;E_{B-V}$.}
\label{fig6}
\end{figure}

\setcounter{table}{2}
\begin{table}
\caption[]{Spectroscopic data for brighter Westerlund 2 members.}
\label{tbl3}
\begin{center}
\begin{tabular}{@{\extracolsep{-2pt}}cccclccc}
\hline \hline
\noalign{\smallskip}
MSP &{\it V} &{\it B--V} &{\it U--B} &Sp.T. &$E_{B-V}$ &$E_{U-B}$ &{\it V--M}$_V$ \\
\noalign{\smallskip}
\hline
\noalign{\smallskip}
18 &12.79 &1.32 &--0.17 &O3~III((f)) &1.65 &1.03 &18.69 \\
32 &15.35 &1.35 &--0.11 &O9.5~V &1.65 &0.99 &18.45$^{\rm a}$ \\
44 &16.00 &1.65 &+0.08 &B1~V &1.92 &1.05 &$\cdots$ \\
151 &14.24 &1.50 &--0.11 &O5~V((f)) &1.83 &1.07 &19.34 \\
157 &14.08 &1.45 &--0.03 &O6~V((f)) &1.78 &1.15 &19.08 \\
167 &14.08 &1.64 &--0.09 &O4~III(f) &1.97 &1.10 &19.88 \\
171 &14.30 &1.65 &+0.03 &O3~V((f)) &1.98 &1.23 &19.70 \\
175 &13.86 &1.33 &--0.03 &O4~V((f)) &1.66 &1.16 &19.06 \\
182 &14.28 &1.46 &--0.07 &O3~V((f)) &1.79 &1.13 &19.68 \\
183 &13.48 &1.70 &+0.05 &O2~V(f*) &2.03 &1.25 &20.08$^{\rm b}$ \\
188 &13.27 &1.49 &--0.06 &O4~III &1.82 &1.13 &19.68 \\
196a &15.77 &1.53 &--0.03 &O8.5~V &1.84 &1.10 &19.27$^{\rm a}$ \\
199 &14.23 &1.55 &+0.02 &O3~V((f)) &1.88 &1.22 &19.63 \\
203 &13.20 &1.52 &--0.06 &O3~V((f))n &1.85 &1.14 &19.10$^{\rm b}$ \\
219 &16.31 &1.45 &+0.33 &O9.5~V &1.75 &$\cdots$ &19.41$^{\rm a}$ \\
229a &15.45 &1.67 &+0.00 &O8~V &1.98 &1.13 &19.85 \\
233 &15.93 &1.59 &--0.07 &O9.5~V &1.89 &1.03 &19.41 \\
235 &16.36 &1.79 &+0.11 &O9.5~V &2.09 &1.21 &20.36 \\
263 &14.79 &1.85 &+0.14 &O3~V((f)) &2.18 &1.34 &20.69$^{\rm b}$ \\
444a &13.20 &1.52 &--0.06 &O4.5~V &1.85 &1.13 &19.55$^{\rm b}$ \\
\noalign{\smallskip}
\hline
\end{tabular}
\end{center}
Notes: $^{\rm a}$ Luminosity for zero-age main sequence adopted.\\
\hspace*{.35in}$^{\rm b}$ Luminosity for class IV adopted.
\end{table}

Color excesses were also calculated for less-reddened stars lying within $1\degr$ of Westerlund~2 using {\it UBV} photometry and spectral types from the literature compiled in Simbad, with the results depicted in Fig.~\ref{fig6}. The data for faint members of Westerlund~2 \citep[mainly from][]{va13} and stars lying near the cluster core exhibit noticeable scatter, most likely because of the effects of crowding by nearby stars on the photometry, and for stars in the surrounding field partly because of variations in the extinction properties of dust affecting individual stars \citep[see][]{tu12}. Overall, there is good agreement with a reddening relation described by $E_{U-B}/E_{B-V}=0.63 + 0.02\;E_{B-V}$, where the reddening slope matches that found by \citet{tu12} for dust affecting the heavily-reddened group of stars associated with WR38, which also belongs to the Great Carina Nebula complex, and the curvature term is adopted from \citet{tu89}. Only 5 of the spectroscopically-observed stars deviate significantly from the adopted relation; that appears to be a consequence of crowding from the close companions for many of the stars.

\begin{figure}
\begin{center}
\includegraphics[width=7cm]{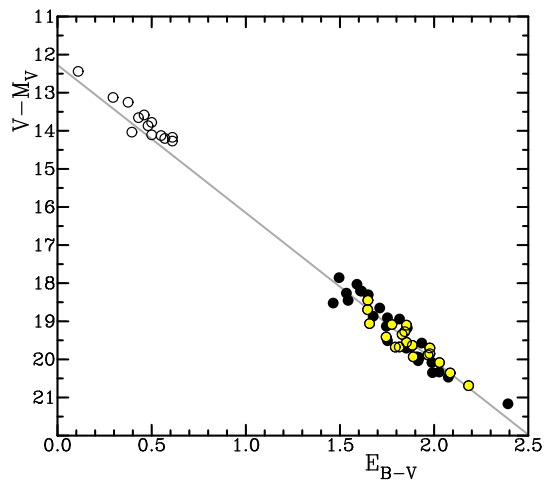}
\end{center}
\caption{A variable-extinction diagram for members of Westerlund~2 from ZAMS fitting (filled circles) and spectroscopic distance moduli (circled yellow points), and for members of IC~2581 (open circles). The gray line represents the best fitting parameters (see text).}
\label{fig7}
\end{figure}

\subsection{Extinction law $R_V$}

{\it UBV} membership assignment in clusters is routinely based on reddening. The {\it UBV} data for Westerlund~2 stars in Table~\ref{tbl3}, and stars without spectroscopic observations identified from dereddening as likely zero-age main-sequence (ZAMS) objects, were used for a variable-extinction study \citep[see][]{jo68} of Westerlund~2. For the former the absolute visual magnitudes are linked closely to the calibration of \citet{wa71b}, while for the latter they were inferred by means of ZAMS fitting techniques \citep[see][]{tu76a,tu76b}, where the ZAMS for O-type stars is that of \citet{tu76a} updated to include the hottest O-type stars with intrinsic colors of {\it (B--V)}$_0$ = --0.33.

Spectroscopically-observed stars in Table~\ref{tbl3} anchored the analysis through their apparent distance moduli, although adjustments were necessary for some stars within their assigned luminosity classifications (see Notes to Table~\ref{tbl3}). Differences of up to $0^{\rm m}.5$ or more exist between the luminosities of dwarfs (class V) and ZAMS stars on one hand, and dwarfs (class V) and subgiants (class IV) on the other, and since it is impossible to distinguish such differences in luminosity class spectroscopically from the available spectra, one must be careful to avoid assigning individual stars erroneous absolute magnitudes that make their parameters appear inconsistent with cluster membership (i.e. reddening, apparent distance modulus). Luminosity adjustments were therefore applied to the few stars indicated above to make them consistent with the general trend of apparent distance modulus versus color excess observed for the large majority of other Westerlund~2 stars. The luminosity calibration used for the spectroscopic absolute magnitudes is tied to the same ZAMS relation \citep{tu80}.

\setcounter{table}{3}
\begin{table}
\caption[]{Recent $R_V$ estimates for the region hosting Westerlund~2.}
\label{tbl4}
\begin{center}
\begin{tabular}{@{\extracolsep{-10pt}}lcll}
\hline \hline
\noalign{\smallskip}
Region &{\it R}$_V$ &Method &Source \\
\noalign{\smallskip}
\hline
\noalign{\smallskip}
West~2 &$3.88\pm0.18$ &{\it UBV}-sp. VE study &This paper \\
IC~2581 &$3.77\pm0.19$ &{\it UBV}-sp. VE study &This paper \\
West~2 &$3.85\pm0.07$ &{\it UBVRIJHK} col.diff. &This paper \\
West~2 &$3.77\pm0.09$ &{\it UBVIJK} sp. fits &\citet{va13} \\
Rup~91 &$3.82\pm0.13$ &{\it UBV} VE study &\citet{te05} \\
Shor~1 &$4.03\pm0.08$ &{\it UBVJHK} CMD fits &\citet{tu12} \\
\noalign{\smallskip}
\hline
\end{tabular}
\end{center}
\end{table}

The resulting analysis for Table~\ref{tbl3} stars in conjunction with ZAMS-fitted stars produced the results depicted in Fig.~\ref{fig7}. Best-fitting relations obtained from least squares and non-parametric techniques restricted to Westerlund~2 stars yielded:
\begin{eqnarray*}
\nonumber
V-M_V & = & V_0-M_V+R_VE_{B-V} \\ \nonumber
& = & 12.27 (\pm0.33) + 3.88 (\pm0.18) E_{B-V} \nonumber
\end{eqnarray*}

\noindent
i.e., $R_V=A_V/E_{B-V}=3.88\pm0.18$ for the ratio of total-to-selective extinction and $V_0-M_V=12.27\pm0.33$ for the intrinsic distance modulus, the latter corresponding to a distance of $2.85\pm0.43$ kpc, closely coincident with the value of 2.8 kpc adopted by \citet{as07} in their independent near-infrared study of Westerlund~2. The large uncertainty in the present distance estimate is the result of the large reddening of cluster stars in conjunction with the extrapolation to $E_{B-V}=0.00$ of the best-fitting relation from a reddening of $E_{B-V}\simeq1.75$.

Such a large value of $R_V$ is consistent with the shallow slope of the reddening line, as also inferred for adjacent regions of Carina \citep{te05,te09,tu12}, and also matches a value of $R_V=3.77\pm0.09$ obtained by \citet{va13} from spectral energy distribution fitting of HST spectra for cluster stars (see Table~\ref{tbl4}).

A similar variable-extinction analysis restricted to stars in Fig.~\ref{fig7} associated with IC~2581 (Turner 1978) yielded values of $R_V=3.77\pm0.19$ and $V_0-M_V=12.01\pm0.09$, the latter corresponding to a distance of $2.52\pm0.11$ kpc. Many young aggregates in the region of the Great Carina Nebula cluster towards distances of $\sim2.1$ kpc \citep{tu12}. Westerlund~2 and IC~2581 appear to lie on the far side of the group.

\begin{figure}
\begin{center}
\includegraphics[width=7cm]{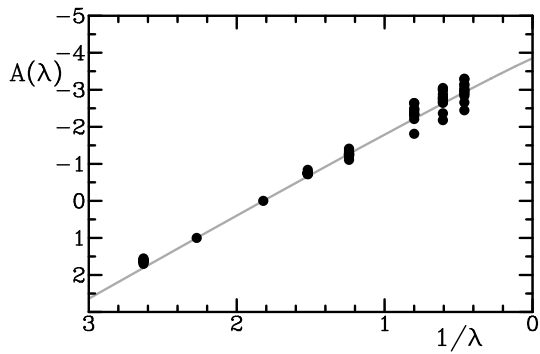}
\end{center}
\caption{Color difference method applied to the {\it UBVR$I_c$JHK}$_s$ magnitudes for Westerlund~2 stars. The fitted relation has $R_V=A_V/E_{B-V}=3.85 \pm0.07$.}
\label{fig8}
\end{figure}

\subsection{Color difference results}

The large value of $R_V$ obtained from the variable-extinction analysis is confirmed by color difference data for Table~\ref{tbl3} stars, and is consistent with other recent studies for the extinction properties of the dust in the region surrounding Westerlund~2, as summarized in Table~\ref{tbl4}. The present use of the color difference method \citep[see][]{jo68} used {\it UBVR$I_c$} data from this study in combination with {\it JHK}$_s$ data from 2MASS \citep{cu03} and intrinsic {\it JHK}$_s$ colours for hot O-stars from \citet{tu11}. The results depicted in Fig.~\ref{fig8} yield an average value of $R_V=3.85\pm0.07$ for 19 spectroscopically-observed stars using the relationship of \citet{fm07} to extrapolate to $\lambda^{-1}=0$ from the infrared region. The curve fitting the data in Fig.~\ref{fig3} is adapted from \citet{zt12} for that value of $R_V$.

\begin{figure}
\begin{center}
\includegraphics[width=7cm]{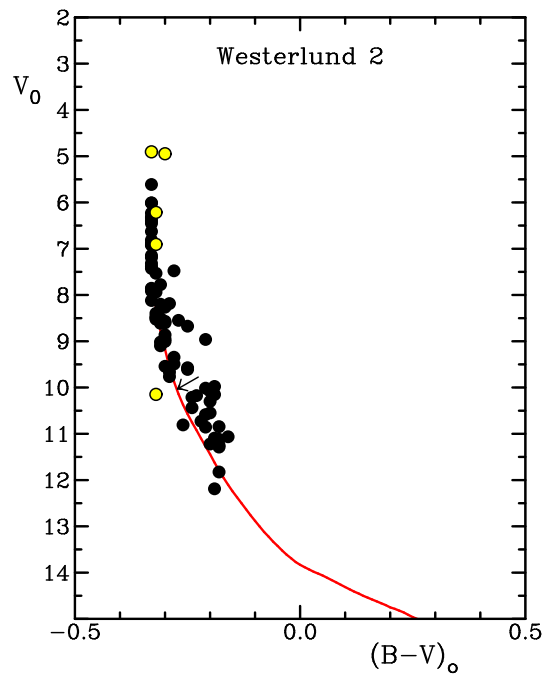}
\end{center}
\caption{Unreddened color-magnitude diagram for Westerlund~2 for the parameters cited in the text, with the Wolf-Rayet stars plotted as circled yellow points, from top to bottom: WR20a, WR20b, WR20c, WR20aa, and WR21a. The red line represents the ZAMS for $V_0-M_V=12.27\pm0.33$, and an arrow depicts the main-sequence gap discussed in the text.}
\label{fig9}
\end{figure}

\section{Westerlund~2}
\label{s-w2}

The unreddened color-magnitude diagram for Westerlund~2 members with good quality {\it UBV} photometry (Fig.~\ref{fig9}) dereddened individually to the intrinsic color-color relation \citep[see][]{tu96,te12b} is that of an extremely young cluster. There are very few hot O-type stars evolved by more than $1^{\rm m}$ from the ZAMS, except for the two Wolf-Rayet stars, WR~20a and WR~20b, which are akin to O-type supergiants. Both have suspected O-supergiant companions \citep{rl11}. Their luminosities are $M_V= -7.37$ and $-7.33$, typical of binary late-WN stars \citep[e.g.,][]{mo93}. According to standard stellar evolutionary models \citep*[e.g.,][]{me93} that implies an age of no more than $\tau \sim2\times10^6$ yr for cluster stars.

Apparent in the cluster color-magnitude diagram (Fig.~\ref{fig9}) is a gap in the main sequence at $V_0\simeq10$. The gap also appears in Fig.~\ref{fig5} as a dearth of reddened cluster members with intrinsic colors of {\it (B--V)}$_0\simeq-0.28$, corresponding to B0.5~V stars. The existence of gaps in the distribution of B-type stars lying on the main sequences in color-magnitude diagrams for young open clusters has been recognized for years \citep[see][]{me82}. \citet{tu96} noted from simulations that gap characteristics are generally consistent with expectations for mergers of close binaries, namely an increased luminosity by $\sim0^{\rm m}.5$ to $3^{\rm m}$ of the former ``main-sequence'' stars from their original position on the cluster main sequence, and a redwards spread in color arising from rapid rotation of the newly-merged systems. That may also be true for Westerlund~2. Such a feature would explain the curious red-wards spread of the cluster main sequence both above and below the gap in Fig.~\ref{fig9}. Presumably the color spread below the B0.5~V gap arises from an additional gap further down the main sequence in the under-sampled region of the cluster color-magnitude diagram.

\subsection{Wolf-Rayet members}

The parameters of the Wolf-Rayet stars near Westerlund~2 are summarized in Table~\ref{tbl5}, where the broad band magnitudes and colors are from the present study, \citet{rl11}, and Simbad, and the spectral types are essentially those of \citet{rl11}. The inferred broad band luminosities are based upon the assumption of membership in Westerlund~2. The close similarity in broad band magnitudes and colors for WR20a and WR20b from this study (Table~\ref{tbl5}) differs markedly from the sizably-different narrow band magnitudes and colors for the same stars \citep{sh91}, and deserves further investigation.

WR20a and WR20b are indicated to be members of Westerlund~2 on the basis of their reddening and location in the core region of the cluster. The assumption of membership in Westerlund~2 also works well for WR20aa and WR20c, despite their location in more distant portions of the cluster corona. Their inferred luminosities are consistent with results for other late-type WN stars in open clusters. \citet{rl11} suggest that they are runaway stars from the cluster. The case for WR21a is much more interesting. Its inferred luminosity as a member of Westerlund~2 is much too small, which suggests that it must lie beyond the cluster. And yet it is also bright with a much smaller reddening than members of Westerlund~2, which would normally imply foreground status to the cluster. The anomaly for this star defies a simple explanation. \citet{rl11} summarize the case for this star as uncertain.

\begin{figure}
\begin{center}
\includegraphics[width=7cm]{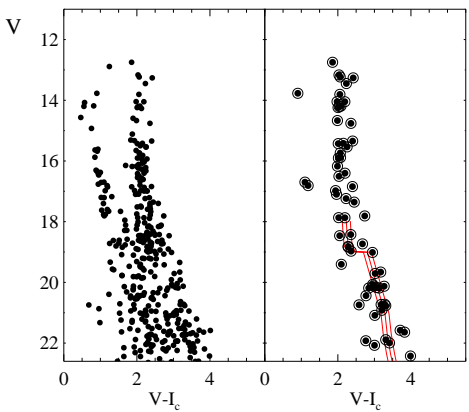}
\end{center}
\caption{Left panel, $VI_c$ data for stars within $1.2 \arcmin$ of the cluster center. Right panel, X-ray (Chandra) cleaned data for that same area reveals the cluster's main-sequence and pre-main-sequence populations. A \citet{si00} $\tau=1.5 \times 10^6$ yr isochrone was applied using the parameters established in \S \ref{s-pa}.}
\label{fig10}
\end{figure}

The dominance of Westerlund~2 members in the X-ray cleaned color-magnitude data of Fig.~\ref{fig10} also suggests that the source of high energy radiation originating from this direction (\S1) must be associated with the cluster stars themselves. Presumably strong stellar winds from the hot stars are responsible for much of the $\gamma$-ray and X-ray flux from this region of Carina.

\setcounter{table}{4}
\begin{table}
\caption[]{Wolf-Rayet stars near Westerlund~2.}
\label{tbl5}
\begin{center}
\begin{tabular}{@{\extracolsep{-4pt}}cccclccc}
\hline \hline
\noalign{\smallskip}
WR &{\it V} &{\it B--V} &{\it U--B} &Sp.T. &{\it (B--V)}$_0$ &$E_{B-V}$ &$M_V$ \\
\noalign{\smallskip}
\hline
\noalign{\smallskip}
20a &13.30 &1.83 &+0.10 &WN6+O3~If* &--0.33 &2.16 &--7.38 \\
20b &13.21 &1.83 &+0.26 &WN6 &--0.30 &2.13 &--7.33 \\
20aa &12.69 &1.17 &$\cdots$ &WN6+O2~If* &--0.32 &1.49 &--5.37 \\
20c &17.51 &2.59 &$\cdots$ &WN6+O2~If* &--0.32 &2.91 &--6.06 \\
21a &12.67 &0.33 &$\cdots$ &WN6+O3~If* &-0.32 &0.65 &--2.13 \\
\noalign{\smallskip}
\hline
\end{tabular}
\end{center}
\end{table}

\section{X-ray Cleaned $VI_c$ CMD}

The cluster age determined from the $UBV$ analysis is corroborated by applying a \citet{si00} isochrone to the $VI_c$ photometry (Fig.~\ref{fig10}). The latter data are deeper and sample the cluster's pre-main-sequence stars. The $VI_c$ photometry was subsequently matched with Chandra X-ray observations \citep{ts07}. The correlation was performed using the cross-match service provided by CDS, Strasbourg. \citet{ts07} note that the X-ray observations were obtained in August 2003 via the Advanced CCD Imaging Spectrometer (ACIS).  ACIS features a $17\arcmin \times 17\arcmin$ field of view, and the cluster was imaged for $40$ ks. The size of the ACIS field is smaller than the optical field (\S \ref{s-ob}), but the cluster core is sampled. Chandra's arc-second resolution is particularly pertinent here given that Westerlund 2 is a compact (crowded) cluster, thus ensuring that sources are reliably matched between the X-ray/optical data.  

The benefits of employing X-ray observations to study stellar clusters are numerous \citep{ev11,ma12}. In this instance the X-ray observations are used to segregate field stars from \textit{bona fide} cluster members, thereby fostering a reliable application of the isochrone. Field stars are typically old slow rotators that have become comparatively X-ray quiet. The X-ray observations highlight those members of Westerlund 2 that feature a late-type companion (chromospherically active), and earlier type members that exhibit prominent stellar winds. The brightest X-ray sources may be linked to colliding mass-loss from the binary WR stars \citep[Table~\ref{tbl5}, see also][]{ts07}.

The left panel in Fig.~\ref{fig10} displays the color-magnitude diagram for all $VI_c$ sources within $1\arcmin.2$ of the cluster center, whereas the right panel contains only X-ray sources occupying that area. Field star contamination is apparent in the former panel, while the cluster's prominence is revealed in the latter. The main-sequence and pre-main-sequence stars are clearly discernible in the X-ray cleaned sample of Fig.~\ref{fig10}, in addition to the gap described in \S \ref{s-w2}. The $VI_c$ X-ray data are well represented by a $\tau=1.5 \times 10^6$ yr isochrone \citep{si00}, which was applied using parameters established in \S \ref{s-pa}. The results confirm the approximate age estimated for Westerlund 2 in \S \ref{s-w2} from the main-sequence turnoff.

\section{Discussion and Conclusions}

Westerlund 2 is found to be $2.85\pm0.43$ kpc distant, in close agreement with the distance to the cluster derived from near-infrared photometry by \citet{as07}. Differences with respect to other estimates can be explained by the anomalous extinction law that applies in the cluster field. In the case of a 1~kpc discrepancy relative to the \citet{va13} study using HST spectra and a similar value of $R_V$, the difference probably originates in the manner of ZAMS fitting, which for \citet{va13} was closer to main-sequence fitting. As noted in \S 3.2, main sequence absolute magnitudes are $\sim0^{\rm m}.5$ more luminous than ZAMS values for O-type stars of identical intrinsic color.

Variable-extinction analyses of members of Westerlund~2 and IC~2581 yield values of $R_V=3.88\pm0.18$ and $3.77\pm0.19$, respectively, while the color difference method applied to 19 Westerlund~2 members observed spectroscopically yields a best estimate of $R_V=3.85\pm0.07$. The agreement of all three estimates is consistent with previous arguments for the existence of anomalous extinction throughout much of the Carina region \citep[][and Table~\ref{tbl4}]{tu12}.

The Wolf-Rayet stars WR20a and WR20b are almost certainly members of Westerlund~2 according to their derived parameters and location in the cluster core. WR20aa and WR20c appear to be members of Westerlund~2 as well, although whether they are runaway cluster members \citep{rl11} or simply members of the cluster corona is left for further study. The cluster color-magnitude diagram (Fig.~\ref{fig9}) displays characteristics that have been noted previously to be consistent with close binary mergers and rapid rotation \citep{tu96}, which, in conjunction with the strong flux of $\gamma$-rays, and X-rays (Fig.~\ref{fig10}) from the cluster reveals Westerlund~2 as an extremely active site for dynamical interactions and mass loss for cluster stars.

The distance established for Westerlund~2 ($\sim 3$ kpc) implies that the cluster lies within the Carina spiral arm.

\begin{acknowledgements}
This study is based on data acquired at Las Campanas Observatory. GC is grateful for the technical support and pleasant environment provided at the observatory. We made use of B. Skiff's Catalogue of Stellar Spectral Classifications, and the WEBDA database maintained in Vienna by E. Paunzen and J.-C. Mermilliod, and the SIMBAD database. GC expresses his gratitude to Edgardo Costa for securing part of the observations discussed in this paper. This study was initiated during a visit by GC to Saint Mary's University in July 2012. ESO funding via the DGDF program for the visit is deeply acknowledged. The scientific results reported here are based in part on observations made by the Chandra X-ray Observatory and published previously by M. Tsujimoto.
\end{acknowledgements}

\end{document}